%% file: Mdot_vs_P.tex
\documentclass[rnote]{aa} 
\usepackage{graphicx}
\usepackage{natbib}
\usepackage[varg]{txfonts}
\usepackage{color}
\usepackage{longtable,lscape}
\begin{document}
\title{Period -- mass-loss rate relation of Miras with and without technetium
  \thanks{Table~\ref{taba1} is only available in electronic form at the CDS via
    anonymous ftp to cdsarc.u-strasbg.fr (130.79.128.5) or via
    http://cdsweb.u-strasbg.fr/cgi-bin/qcat?J/A+A/.}}
\author{S.\ Uttenthaler\inst{1}}
\institute{
  University of Vienna, Department of Astrophysics, T\"urkenschanzstra\ss e 17,
  1180 Vienna, Austria\\
  \email{stefan.uttenthaler@univie.ac.at}
}
\date{Received January 30, 2013; accepted June 07, 2013}

 
\abstract
    {}
    {We report the discovery that Mira variables with and without absorption
      lines of the element technetium (Tc) occupy two different regions in a
      diagram of near- to mid-infrared colour versus pulsation period. Tc is an
      indicator of a recent or ongoing mixing event called the third dredge-up
      (3DUP), and the near- to mid-IR colour, such as the $(K-[22])$ colour
      where [22] is the the 22\,$\mu$m band of the WISE space observatory,
      is an indicator of the dust mass-loss rate of a star.}
   {We collected data from the literature about the Tc content, pulsation
     period, and near- and mid-infrared magnitudes of more than 190
     variable stars on the asymptotic giant branch (AGB) to which Miras belong.
     The sample is naturally biased towards optical AGB stars, which have low to
     intermediate (dust) mass-loss rates.}
   {We show that a clear relation between dust mass-loss rate and pulsation
     period exists if a distinction is made between Tc-poor and Tc-rich Miras.
     Surprisingly, at a given period, Tc-poor Miras are redder in $(K-[22])$
     than are Tc-rich Miras; i.e.\ they have higher mass-loss rates than the
     Tc-rich Miras. A few stars deviate from this trend; physical explanations
     are given for these exceptions, such as binarity or high mass.}
   {We put forward two hypotheses to explain this dichotomy and conclude that
     the two sequences formed by Tc-poor and Tc-rich Miras are probably due to
     the different masses of the two groups. The pulsation period has a strong
     correlation with the dust-mass loss rate, indicating that the pulsations
     are indeed triggering a dust-driven wind. The location in the $(K-[22])$
     vs.\ period diagram can be used to distinguish between pre- and post-3DUP
     Miras, which we apply to a sample of Galactic bulge AGB stars. We find that
     3DUP is probably not common in AGB stars in the inner bulge.}
   \maketitle

\section{Introduction}\label{intro}

Miras are long-period variables in the asymptotic giant branch (AGB) phase of
evolution of low- to intermediate mass stars. On the AGB, rich nucleosynthesis
processes build up heavy elements in the deep interior of the star, most notably
carbon and elements produced by the slow neutron-capture process
\citep[s-process, e.g.][]{Busso99}. One of the elements produced by the
s-process is technetium (Tc), which only has radioactively unstable isotopes.
The products of nuclear burning are brought to the surface of the star by a deep
mixing event called the third dredge-up (3DUP), which may happen after a
powerful ignition of the He-burning shell, called a thermal pulse (TP). As a
result, this phase is also referred to as the TP-AGB. The dredge-up of C may
eventually turn an initially oxygen-rich (M-type, ${\rm C/O}<1$) into a carbon
star (C-type, ${\rm C/O}>1$). Not every TP will be followed by a 3DUP mixing
event; in particular, the first few (weak) TPs are thought to not be followed by
3DUP. It is thought that all powerful pulses on the upper AGB are followed by a
3DUP event, though current models require a parameterization to describe the
mixing process \citep[e.g.][]{Cri11}. Lines of Tc, observable between
4200 -- 4300\,\AA, can be used as a safe indicator of recent or ongoing 3DUP
\citep[e.g.][]{Utt07}. It is important to know which stars undergo 3DUP,
because only these can supply the interstellar matter with heavy elements.

The envelope of a Mira pulsates with a period of 100 to 1000 days, thereby
causing the light variability. The pulsations are thought to play a crucial role
in the mass-loss process from AGB stars because they lift dense atmospheric
material to distances where temperatures are low enough to form dust grains that
are pushed away from the star by radiation pressure, thereby dragging the gas
along. This is a well-understood process in C-stars
\citep[e.g.][]{HD97,Win00,Wac02,Now10,Now11}, but not in the oxygen-rich case
\citep{Woi06}. As recently proposed, light scattering off large transparent
silicate grains may be the driving mechanism in M-type giants
\citep{Hoe08,BH12,Nor12}. The correlation between mass-loss rate and pulsation
period is not very clear in observed samples \citep[e.g.][]{Jura93}, becomes
apparent only when Miras with very long pulsation periods ($P\gtrsim600$\,d) are
included \citep{VW93,Gro09}, or when the selection is restricted to C-rich Miras
\citep{Whi06}.

Near- to mid-infrared colours are a frequently used indicator for the {\em dust}
mass-loss rate ($\dot{M}_{\rm dust}$) of AGB stars. Traditionally, the $(K-[12])$
colour, where $[12]$ is the magnitude in the IRAS 12\,$\mu$m band, has been used
extensively \citep[e.g.][]{Whi94,Whi06}. This type of colour is sensitive to the
dust mass-loss rate because the near-IR (e.g.\ the K-band) probes the
photosphere of the star, while in the mid-IR (e.g.\ [12]) a lot of the light is
emitted by the dusty envelope of the star. The {\em total} mass loss of a star
is, however, dominated by the gas mass-loss rate $\dot{M}_{\rm gas}$, which is
related to $\dot{M}_{\rm dust}$ via the gas-to-dust ratio $\delta$.

In this paper we present the discovery that Tc-poor and Tc-rich Miras fall in
two different regions in a diagram of near- to mid-IR colour (i.e.\
$\dot{M}_{\rm dust}$) versus pulsation period. In both regions, the stars follow a
clear trend towards increasingly red colour with increasing period. The
separation in the two regions is particularly clear when using the K-band
magnitude for the near-IR and the 22\,$\mu$m band of WISE ([22]) as mid-IR
photometry source.

The paper is structured as follows. Section~\ref{data} describes the data
compiled from the literature, Sect.~\ref{res} presents the results, in
Sect.~\ref{disc} an attempt is made to interpret these observations and an
application to bulge samples is introduced, and finally conclusions are drawn in
Sect.~\ref{concl}.

\section{The data}\label{data}

Information on the Tc content was compiled from the following sources:
\citet{Lit87}, \citet{SL88}, \citet{Van91}, \citet{LH99}, \citet{VEJ99},
\citet{Eck00}, \citet{LH03}, \citet{Utt07}, \citet{UL10}, \citet{Utt11}, and
\citet{Smo12}. Further data on the Tc content of carbon stars was taken from
\citet{Abia02} and Barnbaum \citep[priv.\ comm., see also][]{BM93}. Only
classifications ``Tc no'' and ``Tc yes'' were taken into account, ``doubtful'',
``possible'', and ``probable'' cases were disregarded. The different literature
sources used data of various resolutions and signal-to-noise ratios (S/N),
though a resolving power of $R\gtrsim20\,000$ is mandatory to safely detect the
Tc lines in the crowded spectra of cool red giants. Several of the sample stars
have been observed for their Tc content two or more times, and the results
generally agree with each other. Nevertheless, it cannot be excluded that there
are individual objects in the current sample that are misclassified with respect
to their Tc content, though that fraction should be very small.

Those S- and C-type stars without a detection of Tc were not included in the
sample. Thus, we reject any stars that owe their enrichment in C and s-process
elements to mass transfer in a binary system and not to intrinsic nuclear
processing and mixing. The present sample is naturally biased towards stars with
low circumstellar extinction because the Tc lines are located in the blue
spectral range, so that stars in the late superwind phase of the AGB are
missing. Also, most sample stars have periods shorter than 500\,d. It is known
that the number density of optically visible Mira variables drops sharply
between $P=425$ and 500\,d \citep{WC77}.

Since Miras also show pronounced photometric variability in the near-IR, we made
certain to use time-averaged magnitudes in these bands whenever available. Bands
that mainly probe the photosphere of the star (i.e.\ the J- and K-bands, but
also COBE/DIRBE mean fluxes in the 1.25 and 2.2\,$\mu$m bands) were collected
from the following sources: \citet{Cat79}, \citet{Fou92}, \citet{Whi94},
\citet{KH94}, \citet{Ker95}, \citet{Ker96}, \citet{Whi00}, \citet{2MASS},
\citet{Whi06}, \citet{Whi08}, and \citet{Pri10}. The J- and K-band photometry
was converted to the 2MASS system using the relations of \citet{Car01}. Mid-IR
photometry from IRAS, Akari \citep{Ishi10}, and WISE \citep{Wri10} was collected
via VizieR\footnote{http://vizier.u-strasbg.fr/viz-bin/VizieR}. In the
following, these will be referred to as [12], [25], and [60] for the IRAS 12,
25, and 60\,$\mu$m bands; [9] and [18] for the Akari 9 and 18\,$\mu$m bands; and
[22] for the WISE 22\,$\mu$m band.

Periods were taken mainly from \citet{TMW05} where available, or were also
collected from VizieR. Preference was given to sources with available light
curves that allowed for a critical evaluation of the period, such as ASAS
\citep{Poj98}. Since some Miras have pulsation periods that change significantly
in time \citep{WZ81,TMW05}, we also analysed visual pho\-to\-me\-try from
the AAVSO database\footnote{www.aavso.org} and determined present-day periods
with the program {\tt Period04} \citep{LB05}.

In total, data of 197 stars were collected. The data are presented in
Table~\ref{taba1} in the on-line Appendix.

\section{Results}\label{res}

\begin{figure}
\centering
\includegraphics[width=\linewidth,bb=93 368 548 700, clip]{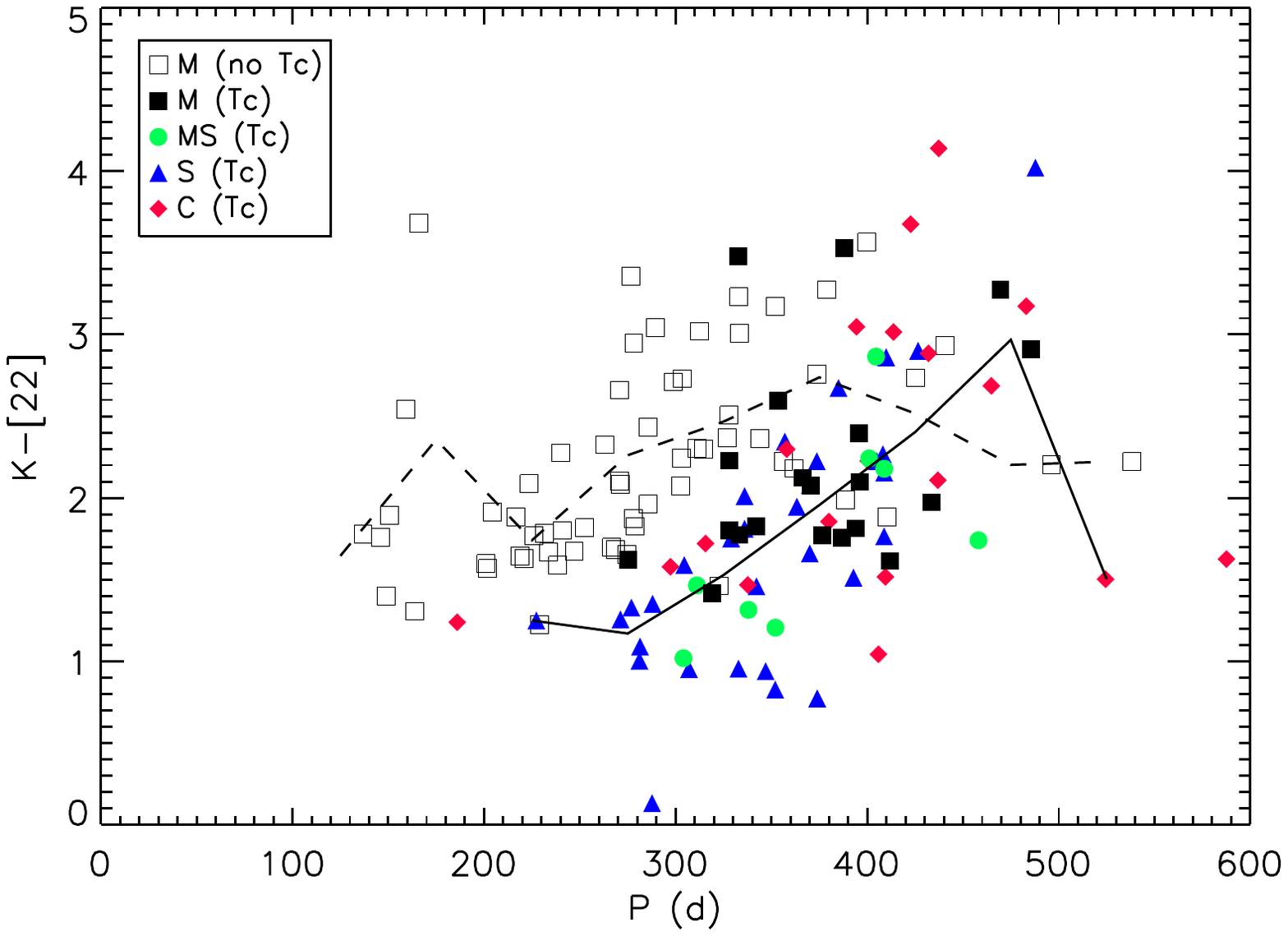}
\caption{Mira stars of a different atmospheric chemistry (spectral type) in the
  $(K-[22])$ vs.\ $P$ diagram. Empty symbols are for Tc-poor stars, filled
  symbols are for Tc-rich stars. The dashed line indicates the run of the mean
  $(K-[22])$ colour in 50\,d period bins for the Tc-poor stars, while the solid
  line shows this run for the Tc-rich stars.}
\label{K22_vs_P}
\end{figure}

Initially, we discovered that Tc-poor and Tc-rich Miras fall into two different
regions in diagrams involving [12], [25], and [60], as well as [9] and [18]
as mid-IR bands. However, it later turned out that the separation is clearest
when using the $(K-[22])$ colour. The WISE catalogue has advantages because of
its all-sky coverage, its relatively high spatial resolution ($12\arcsec$ in the
22\,$\mu$m band), hence little source confusion, and its high sensitivity.
The $(K-[22])$ vs.\ pulsation period diagram of the Miras is presented in
Fig.~\ref{K22_vs_P} and discussed in the following. Also Tc-rich semi-regular
variables (SRVs) form a sequence of increasing $(K-[22])$ colour at short
periods ($P\lesssim200$\,d), but since their periods are less well-defined than
those of the Miras we do not discuss them in detail here.

As can be seen from Fig.~\ref{K22_vs_P}, the Tc-poor stars are {\em redder} in
$(K-[22])$ than the Tc-rich stars over a wide range in period. The run of the
mean colour in 50\,d period bins of the Tc-poor and Tc-rich stars shows that the
difference is $\sim1\fm0$ between $\sim275$ and $\sim375$ days of period. The
distribution of stars is counter-intuitive because one might naively expect the
Tc-rich stars to be more evolved than the Tc-poor ones, thus to have longer
pulsation periods and higher dust mass-loss rates (redder $(K-[22])$ colours).
An interpretation of the distribution is attempted in Sect.~\ref{disc}.

In the meantime, we can notice several facts from this diagram. The two regions
occupied by the Tc-poor and Tc-rich stars follow a trend towards increasing
mass-loss rate with increasing period. This contrasts with previous results
where no such clear trend was found \citep{Jura93}. At $\sim400$\,d period, the
two regions seem to be merging. Interestingly, all Tc-rich stars share roughly
the same region in the diagram, regardless of being M-type (no ZrO bands), MS-
(weak ZrO), S- (strong ZrO), or C-type (C$_2$ and CN bands) stars.

There are two Tc-rich M-type stars (at $P=333$\,d and 388\,d) and one S-type
star (at $P=488$\,d) that have a much redder $(K-[22])$ colour than other
Tc-rich stars of the same spectral type and comparable periods. These deviating
stars are \object{$o$ Cet}, \object{R Aqr}, and \object{W Aql}, all of which
have close binary companions. The dusty environment of R~Aqr and W~Aql as
observed in the far-IR is discussed in \citet{Mayer13}. It is possible that
owing to the binary motion and/or tidal interaction, the companion causes an
enhanced mass-loss rate from the AGB primary stars.

The star \object{R Cet} is a red, Tc-poor outlier (at $P=166$\,d). Regarding the
possibility that binary AGB stars suffer enhanced mass loss, we suggest that
R~Cet is also a member of a binary system. No information on a possible binary
nature of R~Cet was found in the literature, so this is somewhat speculative.

There are also Tc-poor, long-period outliers in the diagram. The two stars at
$P=496$\,d and 538\,d are \object{R Nor} and \object{R Cen}. They are suspected
of being intermediate-mass stars ($M\gtrsim4M_{\sun}$) undergoing hot bottom
burning \citep[HBB; e.g.][]{Boo95}, as signaled by their large lithium abundance
\citep{Utt11,GH13}. It was suggested by observations that the 3DUP efficiency is
low in these stars \citep{GH07}, at least as long as the stars are on the early
AGB with low mass-loss rates. As mentioned, highly obscured stars are missing in
the sample because the Tc lines cannot be investigated for them. Other putative
intermediate-mass, Tc-poor stars, such as \object{W Hya} at $P=389$\,d, may be
the cause of a slight dilution of the otherwise clear separation of Tc-poor from
Tc-rich Miras.

There are also two long-period C-type stars with relatively blue $(K-[22])$
colours in Fig.~\ref{K22_vs_P} (at $P=524$\,d and 588\,d); these are
\object{BH Cru} and \object{LX Cyg}. They have changed their spectral type from
SC to C only in the past decades, at the same time undergoing a strong increase
in pulsation period. They are thus candidates for a recent 3DUP event, and so
their mass-loss rates are expected to rise in the next centuries. Two
forthcoming papers will be dedicated to these stars (Uttenthaler et al., in
preparation).

%
%
%
%

\section{Discussion}\label{disc}

\subsection{Attempt of interpretation}\label{interp}

How can we interpret the two sequences? How do stars evolve in this diagram?
We find from our data that Tc-poor and Tc-rich M- and MS-type stars are
indistinguishable in their $J-K$ colour, so the reason for the separation in
the $(K-[22])$ colour cannot be any effect of increased molecular absorption in
the K-band (and thus relatively fainter $K$ magnitudes). The reasons must
therefore lie in the mid-IR bands. We propose two hypotheses.

Hypothesis 1: Third dredge-up events cause an increase in pulsation period
and/or change the mass-loss properties: for example decrease the total mass-loss
rate, increase the gas-to-dust ratio, and/or decrease the wind velocity of
Miras. In this scenario, the stars start out as Tc-poor Miras, evolve towards
longer period and redder $(K-[22])$ colour until at some point they undergo a
3DUP event. The 3DUP brings up Tc but also C, thereby decreasing the free O in
the atmosphere. The decreased amount of free O (most of it will be locked up in
the CO molecule) changes the (molecular) opacities in the atmosphere that may
lead to an increase in period and/or that changes the mineralogy and amount of
the dust grains in a way that leads to less redistribution of stellar radiation
from the near- to the mid-IR (bluer $(K-[22])$ colour). After dredge-up, the
star continues to evolve to longer period and higher mass-loss rate along the
sequence of the Tc-rich Miras. In this scenario, the stars evolve along zigzag
paths in this diagram.

Hypothesis 2: The two sequences are the result of different masses of the
Tc-poor and Tc-rich stars. The stars that undergo 3DUP already do so before they
start pulsating in the fundamental mode (Mira), that is, they may be irregular
variables or SRVs when they undergo 3DUP, and only later become a Mira
to show up in this diagram. They start out with a longer pulsation period when
they become Miras, at a still relatively low mass-loss rate. Stars with lower
masses, which are those that will never undergo a 3DUP event or do so only very
late in their evolution, start out with shorter periods, but will have a higher
mass-loss rate at a given period than do Tc-rich stars. The lower mass results
in lower surface gravity, which eases mass loss.

We favour Hypothesis~2 for the following reasons. The two sequences are visible
when using different sources of mid-IR photometry, from Akari [9] to IRAS [60].
This would mean that no individual dust feature causes the dichotomy, but rather
a broad difference in near- to mid-IR flux between Tc-poor and Tc-rich stars,
respectively. Thus, it could only be caused by featureless dust species.
It has been suggested that amorphous carbon and metallic iron \citep{McD10} are
present in the envelopes of C- and O-rich AGB stars, respectively. However, it
is not immediately clear why a decrease in free O by 3DUP should lead to
decreased formation of metallic iron grains (i.e.\ increased gas-to-dust ratio),
thereby decreasing the mid-IR flux. Also, it would not be clear why Tc-rich
M-type stars and C-type stars would then occupy roughly the same region in this
diagram. On the other hand, an increase in the pulsation period by changing
molecular opacities is only expected at the ${\rm C/O}<1\rightarrow{\rm C/O}>1$
transition from M- to C-type stars \citep{LW07}. Furthermore, the difference in
C/O between Tc-poor and Tc-rich M-type stars is expected to be small because the
only difference seen in their spectra is in the Tc lines; Tc-rich M-type stars
do not show any signs of ZrO bands in their spectra, which would indicate an
enhanced C/O ratio. It is not clear how a small difference in C/O could lead to
such a strong decrease in the mid-IR flux. After all, the stars are not expected
to be formed with exactly the same C/O ratio, thus a little enhanced primordial
C/O ratio would already shift a Tc-poor star into the Tc-rich regime. This is,
however, not observed. Finally, the Tc-poor, intermediate-mass stars (e.g.\
\object{R Nor} and \object{R Cen}) may occupy another region at an even longer
pulsation period, such that we have a sequence of increasing mass with
increasing period, at a given mass-loss rate. In conclusion, it is unclear how a
3DUP event and the connected change of the chemistry could affect the pulsation
period or the global mass-loss properties of a Mira variable.

The only drawback of Hypothesis~2 may be that there are no Tc-poor Miras
observed at the base of the Tc-rich sequence. Some stars may be expected to
undergo 3DUP only once they are in the Mira pulsation mode. It is possible,
though, that 3DUP, hence Tc enrichment, start early on the AGB, as suggested by
\citet{Jor93}, before Mira pulsation starts.

A test for Hypothesis 2 is the average distance of the stars from the Galactic
plane. More massive, younger stars are expected to have a smaller Galactic scale
height than less massive, older stars. To derive the distance to the stars, we
applied the period-magnitude relation for fundamental-mode pulsators of
\citet[][relation for all stars in Sequence~1 in their Table~6]{Rie10}, assuming
a distance modulus to the Large Magellanic Cloud of 18\fm50. With this approach
we find that the 49 Tc-poor, M-type Miras (excluding the bulge stars and the
putative intermediate-mass stars R~Nor and R~Cen) have a mean absolute distance
from the plane $\left<|Z|\right>=600$\,pc (median 431\,pc), whereas the 74
Tc-rich Miras (all spectral types, excluding bulge stars) have
$\left<|Z|\right>=435$\,pc (median 308\,pc).
Even if selection biases may plague this comparison, this result clearly argues
in favour of the idea that we are dealing with two groups of different average
masses. Different mass is a necessary but not sufficient condition for
Hypothesis~2 to be correct. The sufficient condition would be that heavier stars
start their Mira phase at longer periods and at lower mass-loss rates than less
massive stars.

\subsection{Application to Galactic bulge AGB stars}\label{appli}

\begin{figure}
\centering
\includegraphics[width=\linewidth,bb=93 369 548 699, clip]{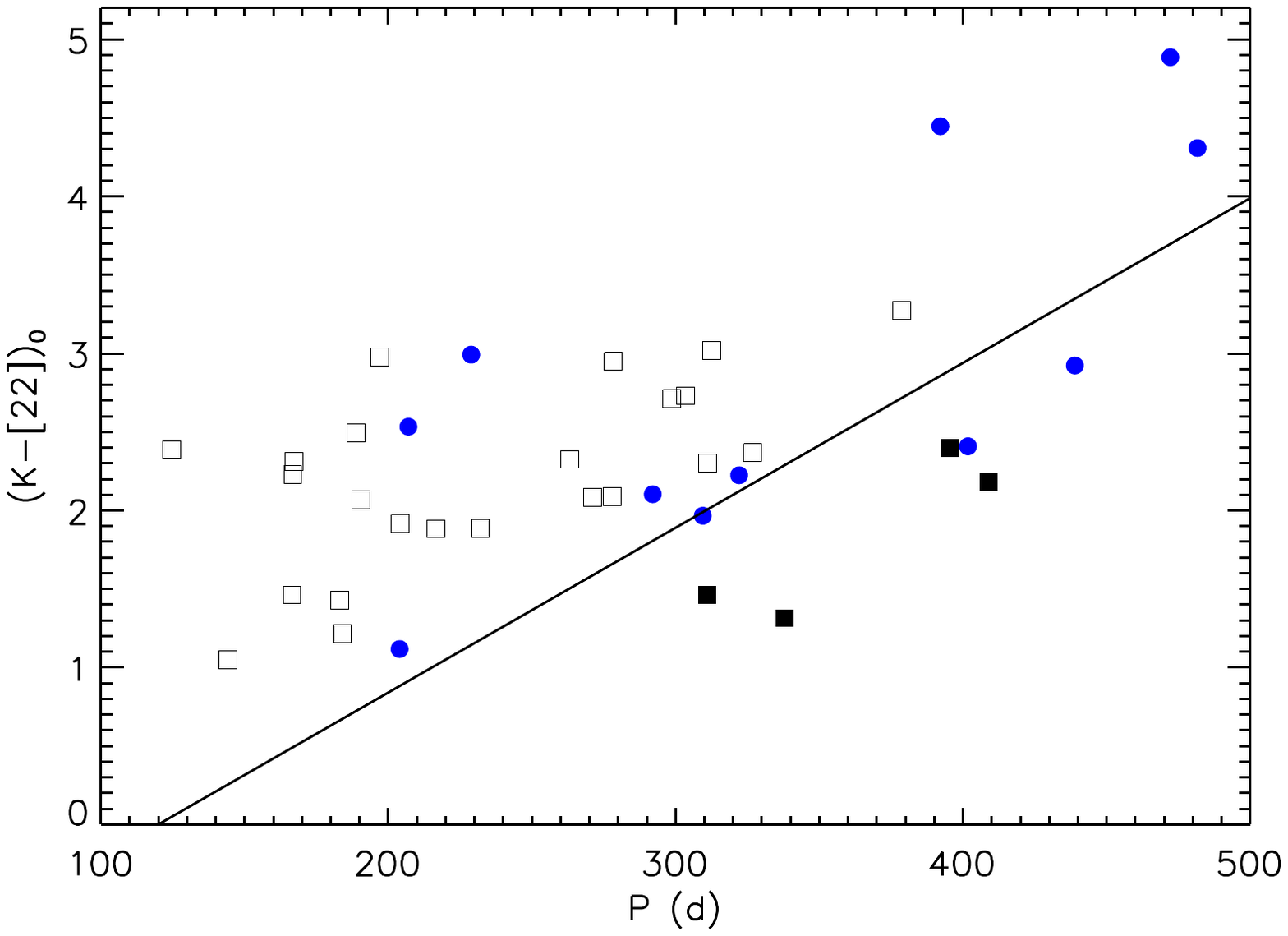}
\caption{$(K-[22])_0$ vs.\ $P$ diagram of Galactic bulge AGB stars. Open and
  filled squares represent Tc-poor and Tc-rich AGB stars in the outer Galactic
  bulge from \citet{Utt07}, whereas filled blue circles represent bulge AGB
  stars from the sample of \citet{Utt13}. The solid line was estimated by eye
  from Fig.~\ref{K22_vs_P} to separate Tc-poor from Tc-rich stars.}
\label{Plaut}
\end{figure}

The separation between Tc-poor and Tc-rich Miras in the $(K-[22])$ vs.\ $P$
diagram is so clear that it may actually be used to distinguish stars that did
undergo a 3DUP event from those that did not. A line to separate pre- from
post-3DUP stars may be drawn between the points $(P,K-[22])=(120,0)$ and
$(P,K-[22])=(520,4.2)$. In Fig.~\ref{K22_vs_P}, 85.4\% of the Tc-poor Miras are
above this line and 87.2\% of the Tc-rich Miras (all spectral types) below this
line. As a result, pre-3DUP Miras can be distinguished from Miras that underwent
a 3DUP event with 85\% confidence. We plot this dividing line in
Fig.~\ref{Plaut}, which includes AGB stars in the outer Galactic bulge (Plaut's
field) that have been investigated for the presence of Tc by \citet{Utt07}.
These stars are also included in Fig.~\ref{K22_vs_P}. The question of 3DUP in
bulge stars is interesting because of the known lack of intrinsic carbon stars
in the Galactic bulge \citep[e.g.][]{Ng97} and the question of mass and age of
bulge stars connected to this. The line separates Tc-poor from Tc-rich stars of
the Plaut sample well. The second group of stars included in Fig.~\ref{Plaut}
are AGB stars from the inner and intermediate bulge from \citet{Utt13} whose
near-IR spectra have been studied at high resolution. No information on the Tc
content of these stars is available. The photometry of these stars has been
de-reddened because they are located in bulge fields with considerable
interstellar extinction. Only stars with $P>125$\,d from that sample are shown
because the shorter period stars are probably SRVs. All of the bulge stars are
O-rich, M-type stars, and only three of the Tc-rich Plaut stars are of type MS
(weak ZrO bands).

Three stars from the inner and intermediate bulge fall below the dividing line
and are thus 3DUP candidates. The one at $P=309$\,d (\object{J174128.5-282733})
was found to have ${\rm C/O}=0.47$ and $^{12}{\rm C}/^{13}{\rm C}=17.4$, but
these values need to be considered as uncertain due to the poor fit achieved to
its near-IR spectrum. Nevertheless, both these ratios are typical of the sample
investigated by \citet{Utt13} and are at the lower limit of what is expected for
a star that underwent 3DUP \citep{SL90}. The other two stars at $P=402$\,d and
439\,d (\object{J175432.0-295326} and \object{J174127.3-282851}) have very
complex near-IR spectra from which it was impossible to derive the C/O or carbon
isotopic ratio. Also, they are in a region in the $(K-[22])$ vs.\ period
diagram where Tc-poor and Tc-rich stars are intermingled. Even if no clear
conclusion on the 3DUP behaviour of these two stars can be drawn, it seems that
3DUP is not wide-spread in this sample of AGB stars in the intermediate and
inner bulge, as also concluded by \citet{Utt13}. It appears that relatively high
metallicity (${\rm [M/H]}\sim-0.2$) and too low a mass
\citep[$M\lesssim1.6M_{\sun}$, see][]{Utt13} act together to prevent efficient
3DUP from forming carbon stars in the Galactic bulge. The Tc-rich Plaut stars in
the outer bulge probably have somewhat lower metallicity \citep{Utt12}, which
would allow for some 3DUP to take place.

\section{Conclusions}\label{concl}

In this paper we showed that Mira variables with and without absorption lines
of technetium occupy two different regions in a diagram of near- to mid-IR
colour ($K-[22]$) versus pulsation period. Both groups of stars follow a trend
towards increasing ($K-[22]$) (dust mass-loss rate) with increasing period. This
clear correlation between pulsation period and mass-loss rate has not been
reported in the literature so far in this period range. Surprisingly, the
Tc-poor stars have redder colours than the Tc-rich stars at any given period,
implying that they have higher (dust) mass-loss rates. We propose two hypotheses
for interpreting this dichotomy. The first one assumes an effect of the third
dredge-up on the pulsation period or mass-loss properties of the stars (total
mass-loss rate, gas-to-dust ratio, wind velocity). The second hypothesis
explains the observed separation as an effect of the different masses of the two
groups of Miras. The Tc-rich Miras would then undergo a dredge-up event early on
the AGB when they are still semi-regular or irregular variables and would only
become Miras later on, with relatively low mass-loss rates. Due to a number of
arguments, we favour the second hypothesis. This conclusion is supported by the
fact that Tc-rich Miras are on average more concentrated towards the Galactic
plane, implying a younger age and a higher average mass of this group. If
Hypothesis~2 is correct, it would mean that at a given pulsation period, more
massive Miras have a lower mass-loss rate.

We note that the scenarios put forward to explain the dichotomy are complex and
involve many details. It is beyond the scope of this paper to investigate the
impact of an increased C/O ratio on the atmospheric structure and dust formation
process or to study the evolution of stars with different masses in the
$(K-[22])$ vs.\ $P$ diagram. Nevertheless, these observations may constitute an
important test of theoretical models of the evolution on the AGB, as well as of
models of dynamic atmospheres and dust formation.

A correlation between pulsation period and $(K-[22])$ colour (i.e.\ dust
mass-loss rate) for Tc-poor and Tc-rich Miras, respectively, clearly supports
the notion that pulsation-enhanced, dust-driven mass loss is also taking place
in oxygen-rich, M-type stars \citep{BH12}.


\begin{acknowledgements}
  The author thanks Cecilia Barnbaum for providing spectra and information on
  the technetium content of carbon stars, and Walter Nowotny and Stefan Meingast
  for careful reading of the manuscript and helpful discussion. The author
  acknowledges support from the Austrian Science Fund (FWF) under project
  P~22911-N16. We acknowledge with thanks the
  variable star observations from the AAVSO International Database contributed
  by observers worldwide and used in this research. This research has made use
  of the SIMBAD database and of the VizieR catalogue access tool, operated at
  CDS, Strasbourg, France. The original description of the VizieR service was
  published in A\&AS 143, 23. This publication makes use of data products from
  the Wide-field Infrared Survey Explorer, which is a joint project of the
  University of California, Los Angeles, and the Jet Propulsion
  Laboratory/California Institute of Technology, funded by the National
  Aeronautics and Space Administration. This publication makes use of data
  products from the Two Micron All Sky Survey, which is a joint project of the
  University of Massachusetts and the Infrared Processing and Analysis
  Center/California Institute of Technology, funded by the National Aeronautics
  and Space Administration and the National Science Foundation.
\end{acknowledgements}

\Online
\begin{appendix}
\include{tablea1}
\end{appendix}

\end{document}

%% file: tablea1.tex
\onllongtab{1}{
\begin{landscape}
\begin{longtable}{lcrcrrcccccccccc}
\caption{The data.}\label{taba1}\\
\hline
\hline
Name & Tc  & Ref. & Period & VarType & SpType & $K$ & $J-K$ & [1.25] & [2.20] & [12] & [25] & [60] &  [9] & [18] & [22] \\
     &     &      & (days) &         &        & mag &       & mag    & mag    & mag  &  mag & mag  & mag  & mag  & mag  \\
 (1) & (2) & (3)  & (4)    & (5)     & (6)    & (7) & (8)   & (9)    & (10)   & (11) & (12) & (13) & (14) & (15) & (16) \\
\hline
\endfirsthead
\caption{Continued.}\\
\hline
\hline
Name & Tc  & Ref. & Period & VarType & SpType & $K$ & $J-K$ & [1.25] & [2.20] & [12] & [25] & [60] &  [9] & [18] & [22] \\
     &     &      & (days) &         &        & mag &       & mag    & mag    & mag  &  mag & mag  & mag  & mag  & mag  \\
 (1) & (2) & (3)  & (4)    & (5)     & (6)    & (7) & (8)   & (9)    & (10)   & (11) & (12) & (13) & (14) & (15) & (16) \\
\hline
\endhead
\hline
\endfoot
\hline
\endlastfoot
\object{R And}       & 1 &1,3 & 409.9 & Mira &  S & $+0.018$ & 1.902 &          &          & $-2.657$ & $-3.493$ & $-3.271$ &          & $-3.145$ & $-2.843$ \\
\object{W And}       & 1 &1,2 & 396.3 & Mira &  M & $+0.431$ & 1.461 &          &          & $-1.927$ & $-2.575$ & $-2.629$ & $-1.076$ & $-1.980$ & $-1.667$ \\
\object{Y And}       & 0 &  1 & 220.9 & Mira &  M & $+4.152$ & 1.101 &          &          & $+2.360$ & $+1.828$ &          & $+3.090$ & $+2.277$ & $+2.522$ \\
\object{Z Ant}       & 1 &  5 & 128.9 &  SRV &  S & $+1.532$ & 1.396 &          &          & $-0.080$ & $-0.813$ & $-0.519$ & $+0.666$ & $-0.526$ & $-0.505$ \\
\object{R Aql}       & 0 &3,10& 270.6 & Mira &  M & $-0.826$ & 1.361 & $+0.538$ & $-0.998$ & $-2.881$ & $-3.903$ &          & $-1.861$ & $-3.182$ & $-3.486$ \\
\object{RR Aql}      & 0 &  7 & 399.8 & Mira &  M & $+0.626$ & 1.576 & $+1.940$ & $+0.311$ & $-2.585$ & $-3.356$ & $-3.369$ & $-1.857$ & $-3.067$ & $-2.939$ \\
\object{RT Aql}      & 0 &  1 & 327.8 & Mira &  M & $+1.120$ & 1.331 &          &          & $-1.051$ & $-1.802$ & $-1.865$ & $-0.356$ & $-1.556$ & $-1.387$ \\
\object{SY Aql}      & 0 &  1 & 356.3 & Mira &  M & $+1.444$ & 1.436 &          &          & $-0.903$ & $-1.713$ & $-1.593$ & $+0.329$ & $-1.100$ & $-0.780$ \\
\object{UV Aql}      & 1 & 12 & 385.5 &  SRV &  C & $+1.558$ & 1.972 & $+3.457$ & $+1.626$ & $+0.100$ & $-0.116$ & $-0.881$ & $+0.587$ & $+0.234$ & $+0.257$ \\
\object{V899 Aql}    & 1 & 11 & 373.7 & Mira &  S & $+5.560$ & 1.440 &          &          & $+3.189$ & $+2.836$ &          & $+3.909$ & $+3.200$ & $+3.336$ \\
\object{V1717 Aql}   & 1 & 11 & 357.0 & Mira &  S & $+5.490$ & 1.540 &          &          & $+3.197$ & $+2.743$ &          & $+3.898$ & $+2.862$ & $+3.146$ \\
\object{W Aql}       & 1 & 11 & 487.8 & Mira &  S & $-0.556$ & 2.090 & $+1.835$ & $-0.114$ & $-4.364$ & $-4.995$ &          &          & $-4.573$ & $-4.578$ \\
\object{R Aqr}       & 1 &  7 & 388.2 & Mira &  M & $-0.821$ & 1.578 & $+0.420$ & $-1.027$ & $-4.367$ & $-4.769$ & $-4.371$ &          & $-3.777$ & $-4.352$ \\
\object{T Aqr}       & 0 &  1 & 201.8 & Mira &  M & $+3.237$ & 1.174 &          &          & $+1.825$ & $+1.438$ & $+1.154$ & $+2.246$ & $+1.583$ & $+1.670$ \\
\object{AU Aur}      & 1 & 13 & 400.5 & Mira &  C & $+2.928$ & 2.308 &          &          & $+0.520$ & $+0.269$ & $-0.058$ & $+0.876$ & $+0.471$ & $+0.702$ \\
\object{AZ Aur}      & 1 & 13 & 413.7 & Mira &  C & $+2.891$ & 2.263 & $+4.086$ & $+2.327$ & $-0.609$ & $-0.818$ & $-1.133$ & $+0.622$ & $-0.246$ & $-0.124$ \\
\object{RU Aur}      & 1 &  7 & 469.6 & Mira &  M & $+0.967$ & 2.073 &          &          & $-1.710$ & $-2.701$ & $-2.472$ &          & $-1.897$ & $-2.309$ \\
\object{SZ Aur}      & 1 & 10 & 458.1 & Mira & MS & $+0.760$ & 1.393 &          &          & $-0.841$ & $-1.407$ & $-1.563$ & $-0.132$ & $-1.263$ & $-0.980$ \\
\object{UV Aur}      & 1 &  1 & 394.4 & Mira &  C & $+2.129$ & 1.900 &          &          & $-0.974$ & $-1.215$ & $-1.130$ &          & $-0.746$ & $-0.918$ \\
\object{VX Aur}      & 0 &  1 & 322.6 & Mira &  M & $+1.527$ & 1.211 &          &          & $+0.130$ & $-0.363$ & $-0.315$ & $+0.715$ & $-0.140$ & $+0.066$ \\
\object{R Boo}       & 0 &  1 & 223.6 & Mira &  M & $+2.106$ & 1.203 &          &          & $+0.491$ & $-0.162$ & $-0.104$ & $+0.746$ & $-0.106$ & $+0.018$ \\
\object{T Cam}       & 1 &  1 & 373.9 & Mira &  S & $+0.814$ & 1.741 & $+1.977$ & $+0.675$ & $-0.410$ & $-0.619$ & $-1.211$ & $+0.158$ & $-0.269$ & $+0.042$ \\
\object{T Cap}       & 0 &  7 & 270.7 & Mira &  M & $+3.297$ & 1.295 &          &          & $+1.349$ & $+0.849$ & $+1.032$ & $+2.023$ & $+1.350$ & $+1.196$ \\
\object{AU Car}      & 1 & 11 & 347.0 & Mira &  S & $+4.270$ & 1.620 &          &          & $+3.320$ & $+3.070$ &          & $+3.743$ & $+3.113$ & $+3.332$ \\
\object{KN Car}      & 1 &  5 & 352.0 & Mira &  S & $+2.516$ & 1.509 &          &          & $+1.668$ & $+1.507$ & $+1.329$ & $+2.159$ & $+1.707$ & $+1.690$ \\
\object{RX Car}      & 1 & 11 & 336.0 & Mira &  S & $+4.900$ & 1.640 &          &          & $+3.066$ & $+2.763$ &          & $+3.779$ & $+3.043$ & $+3.089$ \\
\object{S Car}       & 0 &  1 & 148.9 & Mira &  M & $+1.859$ & 1.354 &          &          & $+0.288$ & $+0.005$ & $-0.153$ & $+0.844$ & $+0.416$ & $+0.462$ \\
\object{U Cas}       & 1 &  1 & 276.9 & Mira &  S & $+2.878$ & 1.451 &          &          & $+1.321$ & $+1.080$ & $+0.872$ & $+1.887$ & $+1.516$ & $+1.550$ \\
\object{W Cas}       & 1 & 13 & 405.9 & Mira &  C & $+2.543$ & 1.765 &          &          & $+1.220$ & $+0.874$ & $+0.222$ & $+1.951$ & $+1.364$ & $+1.501$ \\
\object{R Cen}       & 0 &  1 & 538.1 & Mira &  M & $-0.644$ & 1.286 & $+0.711$ & $-0.555$ &          & $-3.677$ & $-3.271$ & $-1.744$ & $-3.010$ & $-2.865$ \\
\object{V744 Cen}    & 0 &4,5 & 167.3 &  SRV &  M & $-0.809$ & 1.271 & $+0.443$ & $-0.700$ & $-2.236$ & $-3.082$ & $-2.868$ & $-1.412$ & $-2.510$ & $-2.575$ \\
\object{VX Cen}      & 1 &  5 & 109.1 &  SRV &  S & $+0.504$ & 1.413 &          &          & $-0.538$ & $-0.808$ & $-3.690$ & $-0.023$ & $-0.509$ & $-0.448$ \\
\object{T Cep}       & 1 &1,3,10&386.6& Mira &  M & $-1.824$ & 1.328 & $-0.385$ & $-1.751$ & $-3.563$ & $-3.996$ & $-3.859$ &          & $-3.738$ & $-3.580$ \\
\object{omi Cet}     & 1 &1,3 & 332.7 & Mira &  M & $-2.420$ & 1.296 & $-1.368$ & $-2.662$ & $-5.592$ & $-6.315$ & $-6.008$ &          &          & $-5.899$ \\
\object{R Cet}       & 0 &  1 & 165.9 & Mira &  M & $+2.553$ & 1.235 & $+4.028$ & $+2.548$ & $-0.030$ & $-1.251$ & $-1.261$ & $+0.411$ & $-0.718$ & $-1.130$ \\
\object{U Cet}       & 0 &7,9 & 233.7 & Mira &  M & $+2.730$ & 1.110 & $+4.146$ & $+2.782$ & $+1.136$ & $+0.716$ & $+0.722$ &          & $+0.892$ & $+1.062$ \\
\object{W Cet}       & 1 &1,9 & 352.1 & Mira & MS & $+2.140$ & 1.230 & $+3.409$ & $+2.181$ & $+0.820$ & $+0.576$ & $+0.538$ & $+1.396$ & $+0.863$ & $+0.935$ \\
\object{CR Cir}      & 0 &  6 & 108.2 &  SRV &  S & $+2.064$ & 1.253 &          &          & $+1.477$ & $+1.385$ &          & $+1.909$ & $+1.659$ & $+1.783$ \\
\object{BZ CMa}      & 1 & 11 & 332.8 & Mira &  S & $+4.960$ & 1.630 &          &          & $+3.942$ & $+3.642$ &          & $+4.092$ & $+3.659$ & $+4.006$ \\
\object{V355 CMa}    & 1 & 11 & 370.0 & Mira &  S & $+4.760$ & 1.460 &          &          & $+3.353$ & $+2.862$ &          &          & $+3.263$ & $+3.101$ \\
\object{R CMi}       & 1 &  1 & 337.7 & Mira &  C & $+2.491$ & 1.543 &          &          & $+0.689$ & $+0.701$ & $+0.601$ & $+1.510$ & $+1.189$ & $+1.024$ \\
\object{U CMi}       & 0 &  1 & 410.3 & Mira &  M & $+2.581$ & 1.265 &          &          & $+1.006$ & $+0.171$ & $+0.278$ & $+1.692$ & $+0.628$ & $+0.697$ \\
\object{V CMi}       & 1 &  1 & 366.3 & Mira &  M & $+1.965$ & 1.249 &          &          & $+0.135$ & $-0.462$ & $-0.558$ & $+0.938$ & $+0.081$ & $-0.158$ \\
\object{XY CMi}      & 1 & 11 & 271.2 & Mira &  S & $+5.370$ & 1.420 &          &          & $+3.898$ &          &          & $+4.557$ & $+4.011$ & $+4.114$ \\
\object{R Cnc}       & 0 &  1 & 362.0 & Mira &  M & $-0.622$ & 1.298 & $+1.008$ & $-0.491$ & $-2.534$ & $-3.024$ & $-2.961$ & $-1.540$ & $-2.753$ & $-2.802$ \\
\object{R Col}       & 1 &  1 & 328.1 & Mira &  M & $+3.066$ & 1.296 &          &          & $+0.959$ & $+0.158$ & $+0.370$ & $+1.400$ & $+0.578$ & $+0.838$ \\
\object{T Col}       & 0 &  1 & 226.1 & Mira &  M & $+1.957$ & 1.151 & $+3.108$ & $+1.868$ & $+0.039$ & $-0.451$ & $-0.215$ & $+0.611$ & $+0.027$ & $+0.190$ \\
\object{V CrB}       & 1 &12,13&358.0 & Mira &  C & $+1.301$ & 2.152 & $+3.737$ & $+1.814$ & $-1.413$ & $-1.700$ & $-1.806$ & $-0.581$ & $-0.962$ & $-0.997$ \\
\object{X CrB}       & 0 &  1 & 240.9 & Mira &  M & $+3.420$ & 1.388 &          &          & $+1.738$ & $+1.239$ &          & $+2.447$ & $+1.435$ & $+1.619$ \\
\object{BH Cru}      & 1 & 10 & 524.4 & Mira &  C & $+1.560$ & 1.644 &          &          & $+0.329$ & $+0.028$ & $-0.335$ & $+0.384$ & $-0.028$ & $+0.058$ \\
\object{chi Cyg}     & 1 &1,3 & 408.2 & Mira &  S & $-1.902$ & 1.863 & $-0.448$ & $-1.971$ & $-4.440$ & $-4.584$ & $-4.578$ &          &          & $-4.168$ \\
\object{AW Cyg}      & 1 & 12 & 213.0 &  SRV &  C & $+2.114$ & 1.947 & $+1.696$ & $+0.416$ & $+0.254$ & $+0.036$ & $-0.692$ & $+0.745$ & $+0.363$ & $+0.514$ \\
\object{LX Cyg}      & 1 & 10 & 587.5 & Mira &  C & $+2.929$ & 2.199 &          &          & $+1.693$ & $+1.528$ &          & $+1.389$ & $+0.900$ & $+1.304$ \\
\object{R Cyg}       & 1 &  1 & 426.6 & Mira &  S & $+0.861$ & 1.390 &          &          & $-1.424$ & $-2.224$ & $-2.500$ & $-0.547$ & $-1.447$ & $-2.039$ \\
\object{U Cyg}       & 1 & 13 & 464.8 & Mira &  C & $+1.174$ & 2.156 &          &          & $-1.494$ & $-1.815$ & $-2.132$ & $-0.752$ & $-1.445$ & $-1.512$ \\
\object{V441 Cyg}    & 1 &  9 & 286.6 &  SRV &  S & $+1.690$ & 1.390 &          &          & $+0.861$ & $+0.579$ &          & $+1.318$ & $+0.834$ & $+0.960$ \\
\object{V460 Cyg}    & 1 & 12 & 180.0 &  SRV &  C & $+0.270$ & 1.532 & $+1.750$ & $+0.291$ & $-1.073$ & $-1.251$ & $-2.219$ & $-0.544$ & $-0.962$ & $-0.923$ \\
\object{W Cyg}       & 1 &3,4,7&132.8 &  SRV &  M & $-1.424$ & 1.193 & $-0.229$ & $-1.402$ & $-2.728$ & $-3.311$ & $-3.147$ & $-1.959$ & $-2.983$ & $-2.965$ \\
\object{WX Cyg}      & 1 & 13 & 409.5 & Mira &  C & $+2.384$ & 2.081 &          &          & $+1.066$ & $+0.698$ &          & $+1.590$ & $+1.020$ & $+0.868$ \\
\object{R Del}       & 0 &  1 & 285.7 & Mira &  M & $+1.932$ & 1.526 &          &          & $-0.102$ & $-0.619$ & $-0.467$ & $+0.723$ & $-0.246$ & $-0.030$ \\
\object{U Del}       & 1 &4,7 & 119.8 &  SRV &  M & $-0.353$ & 1.223 & $+1.009$ & $-0.161$ & $-1.751$ & $-2.662$ & $-2.395$ & $-1.047$ & $-2.185$ & $-2.324$ \\
\object{Z Del}       & 1 &  1 & 304.5 & Mira &  S & $+4.135$ & 1.154 &          &          & $+2.398$ & $+2.102$ &          & $+2.897$ & $+2.338$ & $+2.547$ \\
\object{R Dor}       & 0 &  4 & 327.1 &  SRV &  M & $-4.227$ & 1.575 & $-2.656$ & $-4.046$ & $-5.652$ & $-5.933$ & $-5.780$ &          &          & $-5.517$ \\
\object{T Dra}       & 1 & 13 & 422.7 & Mira &  C & $+1.637$ & 2.855 &          &          & $-2.107$ & $-2.480$ & $-2.808$ & $-1.622$ & $-2.257$ & $-2.039$ \\
\object{UX Dra}      & 1 &12,13&177.0 &  SRV &  C & $+0.374$ & 1.640 & $+1.747$ & $+0.278$ & $-0.926$ & $-1.149$ & $-1.501$ & $-0.414$ & $-0.726$ & $-0.768$ \\
\object{W Dra}       & 0 & 10 & 289.6 & Mira &  M & $+5.688$ & 1.340 &          &          & $+3.264$ & $+3.022$ & $+2.528$ & $+3.362$ & $+2.666$ & $+2.644$ \\
\object{WZ Dra}      & 1 &  7 & 411.9 & Mira &  M & $+2.764$ & 1.379 & $+4.028$ & $+2.849$ & $+1.249$ & $+0.647$ & $+0.836$ & $+2.172$ & $+1.103$ & $+1.151$ \\
\object{T Eri}       & 0 &  1 & 252.4 & Mira &  M & $+2.444$ & 1.194 & $+4.028$ & $+2.621$ & $+0.749$ & $+0.325$ & $+0.608$ & $+1.361$ & $+0.694$ & $+0.628$ \\
\object{U Eri}       & 0 &  1 & 274.4 & Mira &  M & $+4.047$ & 1.236 &          &          & $+2.588$ & $+2.187$ &          & $+3.183$ & $+2.638$ & $+2.394$ \\
\object{W Eri}       & 0 &  9 & 373.6 & Mira &  M & $+1.780$ & 1.420 & $+2.680$ & $+1.250$ & $-1.340$ & $-1.996$ & $-1.744$ & $-0.179$ & $-0.982$ & $-0.979$ \\
\object{NZ Gem}      & 0 &  2 &  30.1 &  SRV &  S & $+0.557$ & 1.031 &          &          & $+0.126$ & $+0.038$ & $+0.076$ & $+0.551$ & $+0.315$ & $+0.521$ \\
\object{R Gem}       & 1 &  1 & 370.1 & Mira &  S & $+1.664$ & 1.071 & $+2.817$ & $+1.634$ & $+0.293$ & $-0.121$ & $-0.734$ & $+0.489$ & $-0.171$ & $-0.415$ \\
\object{T Gem}       & 1 &  1 & 287.7 & Mira &  S & $+2.705$ & 1.412 &          &          & $+2.200$ & $+2.049$ &          & $+2.668$ & $+2.280$ & $+2.575$ \\
\object{V Gem}       & 1 &  1 & 275.1 & Mira &  M & $+2.808$ & 1.318 & $+4.313$ & $+2.946$ & $+1.215$ & $+0.770$ & $+1.292$ & $+1.876$ & $+1.154$ & $+1.188$ \\
\object{VX Gem}      & 1 &12,13&380.0 & Mira &  C & $+3.130$ & 1.830 & $+5.148$ & $+3.565$ & $+1.046$ & $+0.964$ & $+0.448$ & $+2.013$ & $+1.367$ & $+1.275$ \\
\object{ZZ Gem}      & 1 & 13 & 315.6 & Mira &  C & $+3.236$ & 2.161 & $+4.700$ & $+2.679$ & $+1.308$ & $+1.161$ & $+0.637$ & $+1.846$ & $+1.515$ & $+1.516$ \\
\object{S Gru}       & 1 &7,9 & 401.0 & Mira & MS & $+0.540$ & 1.220 & $+2.106$ & $+0.691$ & $-1.639$ & $-2.331$ & $-2.157$ & $-0.667$ & $-1.805$ & $-1.701$ \\
\object{RU Her}      & 1 &  7 & 485.6 & Mira &  M & $+0.192$ & 1.618 & $+1.652$ & $+0.165$ & $-1.966$ & $-2.662$ & $-2.453$ & $-1.077$ & $-2.138$ & $-2.717$ \\
\object{S Her}       & 1 & 10 & 304.1 & Mira & MS & $+1.044$ & 1.166 & $+2.680$ & $+1.422$ & $-0.209$ & $-0.582$ & $-0.519$ & $+0.521$ & $-0.120$ & $+0.026$ \\
\object{SV Her}      & 0 &  1 & 238.4 & Mira &  M & $+4.707$ & 1.131 &          &          & $+2.711$ & $+2.343$ &          & $+3.463$ & $+2.661$ & $+3.120$ \\
\object{T Her}       & 0 & 10 & 163.9 & Mira &  M & $+2.988$ & 1.222 &          &          & $+1.529$ & $+1.045$ & $+0.993$ & $+2.227$ & $+1.553$ & $+1.682$ \\
\object{R Hor}       & 1 &7,9 & 404.6 & Mira & MS & $-0.880$ & 1.240 & $+0.378$ & $-1.044$ & $-3.526$ & $-4.162$ & $-4.134$ & $-1.567$ & $-3.502$ & $-3.745$ \\
\object{T Hor}       & 0 &7,9 & 218.7 & Mira &  M & $+3.340$ & 1.130 &          &          & $+1.784$ & $+1.373$ & $+1.668$ & $+2.325$ & $+1.601$ & $+1.695$ \\
\object{TW Hor}      & 1 &7,9 & 269.6 &  SRV &  C & $+0.190$ & 1.460 & $+1.493$ & $+0.163$ & $-1.303$ & $-1.836$ & $-1.962$ & $-0.627$ & $-1.347$ & $-1.451$ \\
\object{CZ Hya}      & 1 & 13 & 432.0 & Mira &  C & $+2.434$ & 2.262 & $+4.996$ & $+2.621$ & $-0.667$ & $-1.166$ & $-1.255$ & $-0.000$ & $-0.367$ & $-0.451$ \\
\object{R Hya}       & 1 &7,10& 376.6 & Mira &  M & $-2.663$ & 1.261 & $-1.135$ & $-2.477$ & $-4.374$ & $-4.850$ & $-4.698$ &          &          & $-4.433$ \\
\object{RR Hya}      & 1 &  1 & 342.1 & Mira &  M & $+3.008$ & 1.253 & $+3.561$ & $+2.074$ & $+1.195$ & $+0.764$ & $+0.939$ & $+1.750$ & $+1.101$ & $+1.182$ \\
\object{RU Hya}      & 0 &7,9 & 333.2 & Mira &  M & $+1.600$ & 1.180 & $+3.182$ & $+1.770$ & $-1.091$ & $-1.908$ & $-1.929$ & $-0.005$ & $-0.878$ & $-1.406$ \\
\object{U Hya}       & 1 &12,13&389.4 &  SRV &  C & $-0.716$ & 1.519 & $+0.805$ & $-0.596$ & $-1.894$ & $-2.626$ & $-3.851$ & $-1.614$ & $-2.152$ & $-2.082$ \\
\object{W Hya}       & 0 &4,10& 388.6 & Mira &  M & $-3.215$ & 1.459 & $-1.747$ & $-3.218$ & $-5.429$ & $-5.619$ & $-5.536$ &          &          & $-5.202$ \\
\object{RX Lac}      & 1 &  7 & 331.4 &  SRV & MS & $-0.124$ & 1.324 & $+1.027$ & $-0.215$ & $-1.356$ & $-1.716$ & $-2.121$ & $-0.732$ & $-1.347$ & $-1.419$ \\
\object{S Lac}       & 0 &  1 & 240.2 & Mira &  M & $+2.418$ & 1.305 & $+3.781$ & $+2.481$ & $+0.309$ & $-0.169$ & $+0.047$ & $+1.103$ & $+0.314$ & $+0.143$ \\
\object{RR Lib}      & 0 &  1 & 277.8 & Mira &  M & $+2.513$ & 1.303 &          &          & $+0.804$ & $+0.369$ & $+0.590$ & $+1.416$ &          & $+0.640$ \\
\object{Y Lib}       & 0 &7,9 & 276.6 & Mira &  M & $+3.160$ & 1.210 &          &          & $+0.997$ & $+0.171$ & $+0.318$ & $+1.339$ & $+0.183$ & $-0.199$ \\
\object{S Lup}       & 1 &  1 & 342.2 & Mira &  S & $+2.924$ & 1.430 &          &          & $+1.402$ & $+1.204$ &          & $+2.318$ & $+1.778$ & $+1.466$ \\
\object{RT Lyn}      & 1 &  7 & 394.1 & Mira &  M & $+2.718$ & 1.613 &          &          & $+0.978$ & $+0.551$ & $+0.601$ & $+1.719$ & $+0.874$ & $+0.905$ \\
\object{U Lyn}       & 0 &  7 & 440.5 & Mira &  M & $+1.533$ & 1.478 & $+2.953$ & $+1.362$ & $-1.474$ & $-2.110$ & $-1.825$ & $-0.166$ & $-1.256$ & $-1.398$ \\
\object{HK Lyr}      & 1 &12,13&186.0 & Mira &  C & $+1.689$ & 2.054 & $+3.221$ & $+1.526$ & $+0.239$ & $-0.035$ & $-0.575$ & $+0.724$ & $+0.326$ & $+0.451$ \\
\object{U Mic}       & 0 &7,9 & 332.9 & Mira &  M & $+1.800$ & 1.220 & $+3.695$ & $+2.040$ & $-0.154$ & $-1.025$ & $-1.275$ & $-0.012$ & $-0.864$ & $-1.431$ \\
\object{CL Mon}      & 1 & 13 & 483.0 & Mira &  C & $+1.940$ & 2.788 &          &          & $-1.503$ & $-1.623$ & $-1.996$ & $-0.848$ & $-1.290$ & $-1.234$ \\
\object{SY Mon}      & 0 &  7 & 425.1 & Mira &  M & $+0.848$ & 1.485 &          &          & $-1.796$ & $-2.561$ & $-2.562$ & $-0.792$ & $-2.025$ & $-1.887$ \\
\object{R Nor}       & 0 & 10 & 496.2 & Mira &  M & $+0.804$ & 1.310 & $+2.747$ & $+1.422$ & $-0.747$ & $-1.641$ & $-1.695$ & $+0.128$ & $-1.061$ & $-1.398$ \\
\object{U Oct}       & 0 &  1 & 302.5 & Mira &  M & $+2.163$ & 1.215 & $+3.561$ & $+2.327$ & $+0.606$ & $+0.079$ & $+0.190$ & $+1.133$ & $+0.357$ & $+0.091$ \\
\object{R Oph}       & 0 &  1 & 303.1 & Mira &  M & $+1.019$ & 1.366 & $+2.633$ & $+1.013$ & $-0.911$ & $-1.339$ & $-1.314$ & $-0.357$ &          & $-1.225$ \\
\object{RY Oph}      & 0 &  1 & 150.5 & Mira &  M & $+2.963$ & 1.240 &          &          & $+1.396$ & $+0.918$ & $+0.617$ & $+2.154$ & $+1.368$ & $+1.071$ \\
\object{V Oph}       & 1 &12,13&297.3 & Mira &  C & $+1.689$ & 2.141 & $+4.146$ & $+1.916$ & $-0.027$ & $-0.234$ & $-0.558$ &          & $+0.300$ & $+0.111$ \\
\object{S Ori}       & 1 & 10 & 433.4 & Mira &  M & $-0.500$ & 1.352 & $+1.326$ & $-0.037$ & $-1.818$ & $-2.442$ & $-2.536$ & $-1.397$ & $-2.458$ & $-2.473$ \\
\object{V1365 Ori}   & 0 &  1 &  62.8 &  SRV &  M & $+0.713$ & 1.151 &          &          & $-0.063$ & $-0.109$ & $-0.237$ &          & $+0.222$ & $+0.324$ \\
\object{W Ori}       & 0 &12,13&204.9 &  SRV &  C & $-0.470$ & 1.752 & $+1.303$ & $-0.299$ & $-2.033$ & $-2.214$ & $-2.692$ & $-1.447$ & $-1.950$ & $-1.868$ \\
\object{NU Pav}      & 0 &  7 &  86.5 &  SRV &  M & $-1.526$ & 1.313 & $-0.263$ & $-1.469$ & $-2.280$ & $-2.352$ & $-2.322$ & $-1.751$ & $-2.106$ & $-1.911$ \\
\object{RZ Peg}      & 1 &  1 & 436.9 & Mira &  C & $+2.127$ & 1.807 &          &          & $+0.640$ & $-0.047$ & $-0.222$ & $+1.223$ & $+0.342$ & $+0.019$ \\
\object{S Peg}       & 0 &  7 & 314.4 & Mira &  M & $+1.478$ & 1.280 & $+2.790$ & $+1.397$ & $-0.370$ & $-0.811$ & $-0.701$ & $+0.214$ & $-0.587$ & $-0.819$ \\
\object{TW Peg}      & 0 &  7 &       &  SRV &  M & $-0.539$ & 1.304 & $+0.656$ & $-0.538$ & $-2.416$ & $-3.385$ & $-3.147$ & $-1.535$ & $-2.734$ & $-2.920$ \\
\object{W Peg}       & 0 &  7 & 343.9 & Mira &  M & $+0.007$ & 1.417 & $+1.479$ & $-0.024$ & $-2.217$ & $-2.863$ & $-2.604$ & $-1.244$ & $-2.337$ & $-2.357$ \\
\object{X Peg}       & 0 &  1 & 200.9 & Mira &  M & $+4.468$ & 1.136 &          &          & $+2.933$ & $+2.503$ &          & $+3.556$ & $+2.811$ & $+2.872$ \\
\object{Z Peg}       & 1 &  1 & 327.8 & Mira &  M & $+1.088$ & 1.310 & $+2.496$ & $+1.135$ & $-0.697$ & $-1.315$ & $-1.117$ & $+0.078$ & $-0.918$ & $-0.712$ \\
\object{UZ Per}      & 0 &  7 & 926.7 &  SRV &  M & $+0.894$ & 1.461 &          &          & $-0.888$ & $-1.962$ & $-1.873$ & $-0.029$ & $-1.381$ & $-1.538$ \\
\object{RX Psc}      & 1 & 11 & 281.4 & Mira &  S & $+5.020$ & 1.290 &          &          & $+3.913$ &          &          & $+4.409$ & $+3.840$ & $+3.931$ \\
\object{Z Psc}       & 1 &12,13&278.9 &  SRV &  C & $+0.865$ & 1.515 & $+2.326$ & $+0.885$ & $-0.180$ & $-0.553$ & $-1.104$ & $+0.148$ & $-0.218$ & $-0.146$ \\
\object{CO Pyx}      & 1 & 11 & 329.0 & Mira &  S & $+5.400$ & 1.330 &          &          & $+3.890$ & $+3.554$ &          & $+4.394$ & $+3.429$ & $+3.649$ \\
\object{Y Scl}       & 0 &  9 & 613.0 &  SRV &  M & $+0.350$ & 1.270 &          &          & $-1.494$ & $-2.113$ & $-1.783$ & $-0.448$ & $-1.706$ & $-1.738$ \\
\object{RR Sco}      & 0 &  1 & 278.9 & Mira &  M & $-0.305$ & 1.207 & $+1.128$ & $-0.314$ & $-2.062$ & $-2.547$ & $-2.509$ & $-1.360$ & $-2.258$ & $-2.131$ \\
\object{RZ Sco}      & 0 &  1 & 159.3 & Mira &  M & $+4.152$ & 1.106 &          &          & $+1.582$ & $+1.161$ & $+1.203$ & $+2.745$ & $+1.888$ & $+1.608$ \\
\object{S Sct}       & 1 & 13 & 149.7 &  SRV &  C & $+0.570$ & 1.860 &          &          & $-0.908$ & $-1.025$ & $-2.230$ & $-0.330$ & $-0.680$ & $-0.505$ \\
\object{DX Ser}      & 0 &  7 & 360.0 &  SRV &  M & $+1.471$ & 1.364 & $+2.790$ & $+1.492$ & $+0.100$ & $-0.451$ & $-0.531$ & $+0.830$ & $+0.062$ & $+0.112$ \\
\object{R Ser}       & 1 &1,3 & 353.6 & Mira &  M & $+0.733$ & 1.323 & $+1.730$ & $+0.375$ & $-2.073$ & $-2.560$ & $-2.453$ & $-0.986$ & $-1.931$ & $-1.864$ \\
\object{Y Ser}       & 0 &  7 & 432.7 &  SRV &  M & $+2.249$ & 1.392 &          &          & $+0.340$ & $-0.563$ & $+0.056$ & $+1.330$ & $+0.013$ & $-0.130$ \\
\object{R Sgr}       & 0 &  1 & 268.6 & Mira &  M & $+2.081$ & 1.199 &          &          & $+0.273$ & $-0.058$ & $+0.136$ &          & $+0.159$ & $+0.396$ \\
\object{RV Sgr}      & 1 &  1 & 318.9 & Mira &  M & $+1.625$ & 1.213 & $+3.208$ & $+1.596$ & $-0.023$ & $-0.451$ & $-0.355$ & $+0.829$ & $-0.042$ & $+0.208$ \\
\object{RX Sgr}      & 1 &  1 & 333.3 & Mira &  M & $+2.999$ & 1.335 &          &          & $+1.282$ & $+0.852$ &          & $+1.870$ &          & $+1.226$ \\
\object{T Sgr}       & 1 &1,3 & 392.8 & Mira &  S & $+1.134$ & 1.542 &          &          & $-0.384$ & $-0.818$ & $-1.327$ & $+0.315$ & $-0.340$ & $-0.377$ \\
\object{V781 Sgr}    & 1 &12,13&208.7 &  SRV &  C & $+1.666$ & 1.834 & $+3.334$ & $+1.757$ & $+0.473$ & $+0.195$ &          & $+1.109$ & $+0.617$ & $+0.628$ \\
\object{X TrA}       & 1 &  1 & 364.2 &  SRV &  C & $-0.613$ & 1.785 & $+1.050$ & $-0.522$ & $-2.129$ & $-2.322$ & $-2.737$ & $-1.461$ & $-2.016$ & $-1.862$ \\
\object{R Tri}       & 0 &  1 & 266.5 & Mira &  M & $+0.966$ & 1.181 &          &          & $-0.798$ & $-1.193$ & $-1.133$ & $-0.228$ & $-0.943$ & $-0.730$ \\
\object{T Tuc}       & 0 &7,9 & 247.1 & Mira &  M & $+2.960$ & 1.180 &          &          & $+1.245$ & $+0.764$ & $+0.632$ & $+2.261$ & $+1.392$ & $+1.285$ \\
\object{RR UMa}      & 0 &  1 & 231.4 & Mira &  M & $+5.182$ & 1.259 &          &          & $+3.211$ & $+2.904$ &          & $+3.717$ & $+2.963$ & $+3.399$ \\
\object{S UMa}       & 1 &1,2 &  227.3 & Mira &  S & $+3.041$ & 1.441 &          &          & $+2.064$ & $+1.705$ & $+1.525$ & $+2.724$ & $+2.071$ & $+1.792$ \\
\object{T UMa}       & 0 &  1 & 285.5 & Mira &  M & $+2.739$ & 1.350 & $+4.177$ & $+2.896$ & $+0.836$ & $+0.235$ & $+0.047$ & $+1.137$ & $+0.389$ & $+0.307$ \\
\object{VW UMa}      & 0 &  7 &  66.0 &  SRV &  M & $+1.899$ & 1.219 & $+2.817$ & $+1.720$ & $+1.016$ & $+0.794$ & $+0.647$ & $+1.630$ & $+1.165$ & $+1.324$ \\
\object{VY UMa}      & 1 &12,13&120.4 &  SRV &  C & $+0.484$ & 1.466 & $+1.977$ & $+0.627$ & $-0.658$ & $-0.795$ & $-1.512$ & $-0.159$ & $-0.484$ & $-0.444$ \\
\object{T UMi}       & 0 & 10 & 229.1 & Mira &  M & $+2.894$ & 1.553 &          &          & $+0.734$ & $+0.239$ & $+0.596$ & $+2.409$ & $+1.489$ & $+1.670$ \\
\object{T UMi}       & 0 & 10 & 113.6 &  SRV &  M & $+2.894$ & 1.553 &          &          & $+0.734$ & $+0.239$ & $+0.596$ & $+2.409$ & $+1.489$ & $+1.670$ \\
\object{HP Vel}      & 1 &  5 & 104.1 &  SRV &  S & $+1.893$ & 1.363 &          &          & $+1.063$ & $+0.842$ & $+0.105$ & $+1.591$ & $+1.118$ & $+1.067$ \\
\object{ER Vir}      & 0 &7,9 &       &  SRV &  M & $+1.440$ & 1.100 & $+2.537$ & $+1.479$ & $+0.820$ & $+0.670$ & $+0.635$ & $+1.254$ & $+1.025$ & $+1.170$ \\
\object{EV Vir}      & 0 &7,9 &       &  SRV &  M & $+1.550$ & 1.150 &          &          & $+0.711$ & $+0.497$ & $+0.687$ & $+1.177$ & $+0.803$ & $+0.920$ \\
\object{R Vir}       & 0 &  1 & 146.0 & Mira &  M & $+2.071$ & 1.117 & $+3.425$ & $+2.193$ & $+0.522$ & $+0.098$ & $+0.126$ & $+1.012$ & $+0.427$ & $+0.311$ \\
\object{RS Vir}      & 0 &1,7,9&352.0 & Mira &  M & $+1.150$ & 1.310 & $+2.672$ & $+1.135$ & $-1.464$ & $-2.466$ & $-2.500$ & $-0.708$ & $-2.187$ & $-2.024$ \\
\object{RU Vir}      & 1 & 13 & 437.3 & Mira &  C & $+1.882$ & 2.901 &          &          & $-2.275$ & $-2.530$ & $-2.653$ & $-1.605$ & $-2.169$ & $-2.258$ \\
\object{S Vir}       & 1 &7,9 & 370.4 & Mira &  M & $+0.300$ & 1.270 & $+1.928$ & $+0.513$ & $-1.696$ & $-2.277$ & $-2.102$ & $-0.576$ & $-1.469$ & $-1.777$ \\
\object{X Vol}       & 1 &  5 & 288.0 & Mira &  S & $+3.243$ & 1.497 &          &          & $+1.790$ & $+1.456$ & $+1.992$ & $+2.552$ & $+1.778$ & $+1.892$ \\
\object{R Vul}       & 0 &  1 & 136.8 & Mira &  M & $+3.241$ & 1.462 &          &          & $+1.551$ & $+1.129$ &          & $+2.270$ & $+1.548$ & $+1.463$ \\
\object{RU Vul}      & 0 & 10 & 108.8 &  SRV &  M & $+4.704$ & 1.223 &          &          & $+2.433$ & $+1.890$ & $+1.828$ & $+2.696$ & $+1.541$ & $+1.416$ \\
\object{CSS 466}     & 1 & 11 & 385.0 & Mira &  S & $+4.600$ & 2.240 &          &          & $+2.643$ & $+1.836$ &          & $+3.378$ & $+2.242$ & $+1.927$ \\
\object{CSS 739}     & 1 & 11 & 336.0 & Mira &  S & $+5.500$ & 1.310 &          &          & $+3.288$ & $+2.751$ &          & $+4.004$ & $+3.202$ & $+3.489$ \\
\object{Hen 4-8}     & 1 &  5 & 170.5 &  SRV &  S & $+5.355$ & 1.151 &          &          & $+3.536$ & $+2.780$ &          & $+4.513$ & $+3.402$ & $+3.289$ \\
\object{Hen 4-19}    & 1 &  5 &  48.2 &  SRV &  S & $+5.433$ & 1.176 &          &          & $+4.709$ &          &          & $+5.129$ & $+4.746$ & $+4.898$ \\
\object{Hen 4-20}    & 1 &  5 & 138.9 &  SRV &  S & $+4.277$ & 1.238 &          &          & $+3.370$ & $+3.150$ &          & $+3.943$ & $+3.372$ & $+3.427$ \\
\object{Hen 4-27}    & 1 & 11 & 363.3 & Mira &  S & $+5.160$ & 1.800 &          &          & $+3.288$ & $+2.803$ &          & $+4.127$ & $+3.508$ & $+3.214$ \\
\object{Hen 4-33}    & 1 & 11 & 405.0 & Mira &  S & $+5.480$ & 1.520 &          &          & $+3.133$ & $+2.506$ &          & $+3.991$ & $+3.078$ & $+3.255$ \\
\object{Hen 4-36}    & 1 &  5 &  56.8 &  SRV &  S & $+2.653$ & 1.397 &          &          & $+1.727$ & $+1.601$ & $+1.360$ & $+2.282$ & $+1.864$ & $+1.974$ \\
\object{Hen 4-37}    & 1 &  5 & 108.6 &  SRV &  S & $+4.273$ & 1.223 &          &          & $+3.361$ & $+3.092$ &          & $+3.896$ & $+3.460$ & $+3.638$ \\
\object{Hen 4-39}    & 1 &  5 &  46.6 &  SRV &  S & $+3.209$ & 1.327 &          &          & $+2.768$ & $+2.700$ &          & $+3.305$ & $+2.940$ & $+3.051$ \\
\object{Hen 4-41}    & 1 &  5 & 125.1 &  SRV &  S & $+2.727$ & 1.626 &          &          & $+1.514$ & $+1.199$ & $+1.157$ & $+2.160$ & $+1.688$ & $+1.756$ \\
\object{Hen 4-64}    & 1 &  5 &  59.1 &  SRV &  S & $+3.495$ & 1.487 &          &          & $+2.584$ & $+2.299$ &          & $+3.046$ & $+2.627$ & $+2.733$ \\
\object{Hen 4-66}    & 1 &  5 & 216.0 &  SRV &  S & $+4.277$ & 1.328 &          &          & $+3.638$ & $+3.395$ &          & $+4.108$ & $+3.614$ & $+3.791$ \\
\object{Hen 4-81}    & 1 & 11 & 408.8 & Mira &  S & $+4.990$ & 1.790 &          &          & $+3.640$ & $+3.141$ &          & $+4.169$ & $+3.504$ & $+3.226$ \\
\object{Hen 4-84}    & 1 & 11 & 408.9 & Mira &  S & $+4.630$ & 1.860 &          &          & $+2.508$ & $+1.909$ &          & $+3.704$ & $+2.764$ & $+2.475$ \\
\object{Hen 4-95}    & 1 &  5 &  83.3 &  SRV &  S & $+3.798$ & 1.084 &          &          & $+2.272$ & $+1.285$ &          & $+2.927$ & $+1.934$ & $+1.663$ \\
\object{Hen 4-104}   & 1 &  5 &  40.3 &  SRV &  S & $+4.160$ & 1.252 &          &          & $+3.526$ & $+3.150$ & $+0.762$ & $+4.045$ & $+3.677$ & $+3.769$ \\
\object{Hen 4-109}   & 1 & 11 & 281.1 & Mira &  S & $+4.830$ & 1.370 &          &          & $+3.788$ & $+3.606$ &          & $+4.345$ & $+3.681$ & $+3.827$ \\
\object{Hen 4-122}   & 1 & 11 & 307.0 & Mira &  S & $+5.260$ & 1.620 &          &          & $+4.031$ &          &          & $+4.596$ & $+4.038$ & $+4.309$ \\
\object{Plaut 3-45}  & 0 &  8 & 271.0 & Mira &  M & $+6.650$ & 1.450 &          &          &          &          &          & $+5.580$ &          & $+4.567$ \\
\object{Plaut 3-70}  & 0 &  8 & 166.5 &  SRV &  M & $+7.150$ & 1.290 &          &          &          &          &          & $+6.346$ &          & $+5.688$ \\
\object{Plaut 3-100} & 0 &  8 & 298.7 & Mira &  M & $+6.370$ & 1.290 &          &          & $+3.879$ & $+2.244$ &          & $+4.958$ & $+3.236$ & $+3.660$ \\
\object{Plaut 3-143} & 0 &  8 & 204.2 & Mira &  M & $+7.960$ & 1.250 &          &          &          &          &          & $+6.817$ &          & $+6.045$ \\
\object{Plaut 3-195} & 0 &  8 & 216.6 & Mira &  M & $+7.580$ & 1.210 &          &          &          &          &          & $+6.102$ &          & $+5.697$ \\
\object{Plaut 3-277} & 0 &  8 & 263.2 & Mira &  M & $+7.110$ & 1.320 &          &          &          &          &          & $+5.835$ &          & $+4.785$ \\
\object{Plaut 3-315} & 0 &  8 & 326.8 & Mira &  M & $+6.540$ & 1.370 &          &          & $+4.788$ &          &          & $+5.142$ & $+4.032$ & $+4.172$ \\
\object{Plaut 3-328} & 0 &  8 & 166.8 &  SRV &  M & $+7.750$ & 1.200 &          &          &          &          &          & $+6.462$ &          & $+5.522$ \\
\object{Plaut 3-331} & 0 &  8 & 311.1 & Mira &  M & $+6.600$ & 1.460 &          &          & $+4.455$ &          &          & $+5.212$ & $+4.143$ & $+4.299$ \\
\object{Plaut 3-626} & 1 &  8 & 311.1 & Mira & MS & $+7.190$ & 1.260 &          &          &          &          &          & $+6.027$ &          & $+5.726$ \\
\object{Plaut 3-639} & 0 &  8 & 167.3 &  SRV &  M & $+7.450$ & 1.260 &          &          &          &          &          & $+6.165$ &          & $+5.136$ \\
\object{Plaut 3-719} & 0 &  8 & 277.8 &  SRV &  M & $+6.610$ & 1.300 &          &          &          &          &          & $+5.382$ & $+4.162$ & $+4.523$ \\
\object{Plaut 3-794} & 0 &  8 & 303.5 & Mira &  M & $+6.040$ & 1.350 &          &          &          &          &          & $+4.613$ & $+3.600$ & $+3.310$ \\
\object{Plaut 3-942} & 1 &  8 & 338.0 & Mira & MS & $+6.550$ & 1.340 &          &          &          &          &          & $+5.888$ &          & $+5.236$ \\
\object{Plaut 3-1002}& 0 &  8 & 190.5 &  SRV &  M & $+7.080$ & 1.220 &          &          &          &          &          &          &          & $+5.012$ \\
\object{Plaut 3-1008}& 0 &  8 & 232.1 &  SRV &  M & $+6.720$ & 1.250 &          &          &          &          &          & $+5.844$ &          & $+4.835$ \\
\object{Plaut 3-1059}& 0 &  8 & 144.1 &  SRV &  M & $+7.690$ & 1.230 &          &          &          &          &          &          &          & $+6.641$ \\
\object{Plaut 3-1147}& 1 &  8 & 395.6 & Mira &  M & $+5.770$ & 1.500 &          &          & $+3.405$ & $+2.763$ &          &          &          & $+3.373$ \\
\object{Plaut 3-1176}& 0 &  8 & 184.1 &  SRV &  M & $+6.940$ & 1.330 &          &          &          &          &          &          &          & $+5.724$ \\
\object{Plaut 3-1179}& 0 &  8 & 278.2 & Mira &  M & $+7.200$ & 1.360 &          &          & $+3.881$ & $+2.961$ &          & $+5.599$ & $+4.253$ & $+4.251$ \\
\object{Plaut 3-1204}& 0 &  8 & 197.0 &  SRV &  M & $+7.310$ & 1.330 &          &          & $+4.807$ & $+2.847$ & $-0.229$ & $+5.519$ & $+4.236$ & $+4.334$ \\
\object{Plaut 3-1287}& 0 &  8 & 312.5 & Mira &  M & $+6.640$ & 1.310 &          &          &          &          &          & $+4.828$ &          & $+3.620$ \\
\object{Plaut 3-1313}& 0 &  8 & 378.7 & Mira &  M & $+6.249$ & 1.470 &          &          & $+3.884$ & $+2.797$ &          & $+4.656$ & $+3.570$ & $+2.974$ \\
\object{Plaut 3-1347}& 1 &  8 & 408.9 & Mira & MS & $+5.990$ & 1.480 &          &          & $+3.983$ & $+3.109$ &          & $+4.429$ & $+3.517$ & $+3.812$ \\
\object{Plaut 3-1470}& 0 &  8 & 183.2 &  SRV &  M & $+7.340$ & 1.210 &          &          &          &          &          & $+6.570$ &          & $+5.911$ \\
\object{Plaut 3-1517}& 0 &  8 & 188.8 &  SRV &  M & $+7.680$ & 1.260 &          &          &          &          &          & $+6.536$ &          & $+5.185$ \\
\object{Plaut 3-1991}& 0 &  8 & 124.7 &  SRV &  M & $+8.020$ & 1.150 &          &          &          &          &          &          &          & $+5.633$ \\
\end{longtable}
\tablefoot{Meaning of the columns: (1): name of the star; (2): Tc content (0=Tc-poor,
            1=Tc-rich); (3): reference for Tc content; (4): pulsation period;
            (5): variability type;
            (6): spectroscopic type; (7): K-band magnitude; (8): $J-K$
            colour; (9) and (10): COBE/DIRBE 1.25 and 2.20\,$\mu$m magnitude
            \citep{Pri10}; (11), (12), and (13): IRAS 12, 25, and 60\,$\mu$m
            magnitude; (14) and (15): Akari 9 and 18\,$\mu$m magnitude;
            (16): WISE 22\,$\mu$m magnitude. References for the Tc content:
            1: \citet{Lit87}, 2: \citet{SL88}, 3: \citet{Van91}, 4: \citet{LH99},
            5: \citet{VEJ99}, 6: \citet{Eck00}, 7: \citet{LH03},
            8: \citet{Utt07}, 9: \citet{UL10}, 10: \citet{Utt11},
            11: \citet{Smo12}, 12: \citet{Abia02}, 13: \citet{BM93}. The
            following zero-magnitude fluxes were used to convert fluxes to
            magnitudes: 1593.7 and 648.3\,Jy for COBE/DIRBE 1.25 and
            2.20\,$\mu$m; 28.3, 6.73, and 1.19\,Jy for IRAS 12, 25, and 60;
            58.08 and 10.77\,Jy for Akari 9 and 18, respectively.
            \object{T UMi} is listed twice because it recently underwent a
            marked period decrease and switched from fundamental mode (Mira)
            to first overtone (SRV) pulsation \citep{Utt11}.}
\end{landscape}
}